# Ion irradiation induced direct damage to proteins and their components

Wei WANG [*†‡]   Zengliang YU‡   Wenhui SU †

[†]Center for the Condensed Matter Science and Technology, Harbin Institute of Technology, Harbin 150080, China

[‡]Key Lab of Ion Beam Bioengineering, Institute of Plasma Physics, Chinese Academy of Sciences, P. O. Box 1126, Hefei, 230031, China

[*]Corresponding author. Email: wwang_ol@hit.edu.cn

**Abstract:** In our last paper, we have reported the direct interact between energetic ions and DNA, and its components, pointing out that damage to DNA is actually of highest biological consequence, as the integrity of DNA sequence is essential for biological function. Proteins are another kind of important biomolecules. Although much recent attention has been payed to ion irradiation initiated reaction cascades leading to DNA damage, the protein lesions induced by energetic ions remain virtually unexplored at the molecular scale (maybe because there are usually many copies of each protein, so a comparable level of damage is of much less biological impact). Except for DNA damage, an understanding of nascent events leading to radiolytic protein damage is also required in order to achieve a complete description of ionizing radiation effects on living cells.

1. Amino acids

Except for DNA damage, an understanding of all nascent events leading to radiolytic protein damage is also required in order to achieve a complete description of ionizing radiation effects on living cells. As the fundamental building blocks of proteins, the radiation-induced free radical formation of solid-state amino acids has been the subject of numerous investigations by ESR, and this subject has been reviewed[1]. The radical species produced in solid-state amino acids by keV ion irradiation seem to be more abundant than by photon irradiation (e.g., X-ray, γ-ray), because there are more fragments due to the direct nuclear collision process[2]. Under different ion energy conditions, ESR spectra of heavy-ion-bombarded samples show different radical species with different individual contribution. Conversions into a deamination radical, •CHR-COO$^-$ and a dehydrogenation one, •CR(NH$_3^+$)-COO$^-$, are the two main channels for radical formation of amino acids. In terms of dose effects, disparate particles such as keV N-ions, MeV protons and helium ions have a similar observed relationship between ion dose and radical yield trends, with a maximal yield value at a medium dose[3,4]. After ion irradiation, the radical concentration declines rapidly as the result of radical quenching. It shows an exponential decay initially and remains stable later, supporting the existence of long-lived radicals[2].

A second point about ion irradiation of amino acids is molecular fragmentation. The integrated effects can be roughly quantified by the sputtering yield. It was reported that each incident ion with energies between 10 and 100eV can break $10^3$ to $10^4$ bonds[5]. High-yield molecular emission of $H_2$, $NH_3$, $CO_2$ and dimer hydrocarbons has been detected during 200keV helium and argon ion bombardment of alanine and glycine[6,7]. The common neutral losses such as $NH_3$, $H_2O$ and COOH have also been characterized as the main channels of amino acid fragmentation using mass spectrometry techniques[8,9]. The molecular ion or quasi-molecular ion is not the main peak anymore due to collision induced fragmentation.

Ion irradiation can also give rise to molecular crosslinking. Several new larger molecules can be formed per incident ion by fragment recombination. GC-MS (Gas Chromatograph-Mass Spectrometer) analysis showed that 30keV nitrogen ion irradiation of glycine lead to the formation of a variety of products, including sputtered species with a higher molecular weight than glycine itself [10,11]. This phenomenon has also been observed in the mass spectrum of glycine induced by FAB (Fast Atom Bombardment)-CID. More interestingly, it was found that oligopeptides are easily produced from amino acid monomers and their mixtures in solid state, initiated by high energy (6.6MeV) protons[12]. Recently we confirmed and extended these results by Simakov: oligopeptides can be synthesized from their monomers by using energetic ions ranging from keV to MeV; the peptides are possibly formed through radical-based condensation reaction[13]. These thoughtprovoking results are of great importance from a broad astrobiological point of view.

Energetic ions can distribute their energy steadily along their tracks and produce displaced atoms when they approach target molecules. If they are active elements, after slowing down to the energy range in which chemical reactions can occur, they may bond with neighboring displaced atoms or matrix atoms to form new chemical species that contain the incident ions.. This is the so-called mass deposition effect. Based on the measured UV, IR (infrared), Raman, and Mass spectra in conjunction with quantum theory, Shao et al[14] deduced the reaction pattern of a 30keV $N^+$ ion with tyrosine (Tyr) molecule and suggested a substitution reaction between the decelerated $N^+$ and the displaced $C_5$ of the Tyr benzene ring (Fig. 1). As a result, a new nitrogen-containing heterocyclic compound has been formed and a carbonyl group is detected from the irradiated Tyr sample. In the case of molecular modification of cysteine by using a 110 keV $Fe^+$ ion beam, ESI-FTMS (electrospray ionization-Fourier transform mass spectrometry) shows the characteristic mass spectra of three possible iron-containing coordination compounds [15]. These interesting results, beyond all doubt, give support to the hypothesis of mass deposition effect of ion irradiation.

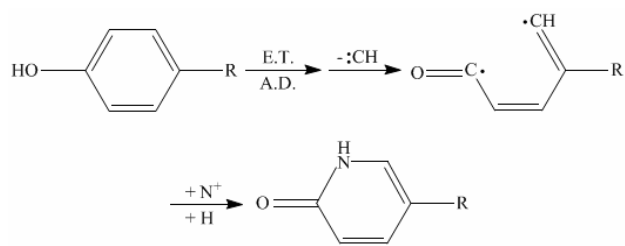

Fig.1 keV N-ion implantation induced nitrogen-containing heterocycle formation from the tyrosine molecule. E.T. = energy transfer; A.D. = atomic displacement; R= -CH$_2$CH(NH$_3^+$)COO$^-$.

2. Peptides

Homopolymers of the amino acids gly, ala, val, phe and tyr have been used as models to investigate the effect of molecular structure on the degradation of polypeptides by ionizing radiation[16]. The main reactions are demonstrated to be: 1) scission of the entire side chain leading to its hydrogen adduct; 2) fragmentation of the side chain; 3) main-chain scission and fragmentation. Among these three channels, none is exclusive, but also none can be rejected. More than one of the three channels may occur during ion irradiation induced peptide fragmentation. The most favorable site of fragmentation may be dictated by the lowest activation barrier. Under low-energy collision-induced dissociation conditions, peptides are suggested to typically fragment at their peptide bonds[17].

To investigate the exact fragmentation pattern of peptides by ion bombardment, mass loss measurement is used to derive the sputtering yield when further details of the mass loss mechanism are not known. The recent developments of multiple-stage mass spectrometry (MS$_n$) have made mass spectrometry the instrument of choice for peptide fragmentation research. Sequence ions that arise by cleavage of the peptide chain can be identified and heterogeneous cleavage is observed from both ends of the chain[18]. Those complexes contain the C-terminal part as a smaller linear peptide and the N-terminal part either as an oxazolone or a cyclic peptide. Biomolecular fragmentation studies with slow HCI show a strong dependence of the positive and negative ion yields on the charge state of the incident ion (Xe$^{n+}$); the ions with high charge states cause the ejection of fragments with a wide mass range as well as the intact molecule [19].

Peptides with special amino acid components usually have special fragmentation strategy. By using Raman, IR, and XPS (X-ray Photoelectron Spectroscopy), Compagnini and co-workers studied the effects of low fluence 250 keV He$^+$ irradiation on tripeptide glutathione (GSH, γ-L-glutamyl-L-cysteinyl-glycine) thin films in their investigation of the protective roles in cells against ionizing radiation [20]. It was observed that the GS• radical was formed and survived for a long time in the solid state. The chemical yield of the reaction GSH → GS• was estimated to be about 1000 (reactions) per impinging ion. The lack of S-S signal (Raman) can be understood in terms of no presence of the oxide form GS-SG. For H, alkyl, and aryl side chain-containing peptides, their CID induced fragmentation processes are very complicated and quite different from each other[21,22].

3. Proteins

Photon-irradiation of proteins generates specific structural and chemical alternations, including fragmentation, decarboxylation, aggregation, breakage of disulfide bonds, the formation of disulfide radicals, the release of aromatic amino acids, loss of helicity, and so on[23-25]. These lesions are expected to be able to occur during ion bombardment of proteins. However, compared with the attention that has been drawn to protein damage induced by photon radiation, there are a few reports about particle irradiation on proteins, possibly because macro-molecules such as proteins cannot be easily prepared for ion irradiation research.

Protein aggregation was often reported in photon irradiation research[24]. But no aggregation was observed in 30keV N$^+$ irradiation on bovine serum albumin (BSA) [26]. FTIR (Fourier Transform Infrared) shows that the regular secondary structures of BSA molecules are very sensitive to ion irradiation. With increasing ion fluences from 0 to $2.5 \times 10^{16}$ ions/cm$^2$, the fractions of α-helix and β-turns decrease from 17 to 12% and from 40 to 31%, respectively; while those of the random coil and β-sheet structure increase from 18 to 27% and from 25 to 30%, respectively. These structural alterations in the proteins may be detrimental to the organism and may result in the mutation or death of cells.

As mentioned before, mass spectrometry of biomolecules is feasible with the development of methods to produce ionic forms of relatively large molecules such as peptides, and to provide some useful information about protein fragmentation[27]. The dissociation behaviors of protonated bovine ubiquitin ions has been studied for charge states ranging from +6 to +12 on a modified triple quadrupole/linear ion trap tandem mass spectrometer[28]. Extensive cleavage along the protein backbone yields richer sequence information. About 77% and 52% of backbone amide bond cleavages happen in the beam-type CID and ion trap CID, respectively. Reversible protein phosphorylation is a ubiquitous cellular regulatory mechanism for the control of signal transduction networks in organisms. FABMS experiments showed that under keV fast atom bombardment conditions, the dominant fragmentation pattern of phosphorylation sites is cleavage of the phosphate ester bond[29].

A well-ordered highly-specific 3D structure is widely accepted as the necessary prerequisite for protein function. Ion irradiation induced protein damage such as fragmentation, dephosphorylation, chain cleavage, and defolding, will inevitably give rise to protein denaturation, loss of function, and even cell apoptosis. On the other side, in biological systems proteins are involved directly or as DNA binding motifs in various cellular activities such as DNA packaging, regulation of gene expression, and enzymatic catalysis. Therefore, the above-mentioned lesions of proteins and their components can not only denature proteins, but may also induce DNA damage and be fatal to organisms. It has been found that unlike the case of X-rays, deviant DNA-protein crosslinks are included in the residual DNA lesions after nitrogen ion irradiation[30]. Moreover, it has also been observed that in a *lac* repressor experimental system in Escherichia coli, argon ion irradiation cleaves the peptide chain of the *lac* repressor protein and the protein loses its *lac* operator-binding activity, and hence the protein-DNA complex is disrupted[31].